\documentstyle[aps]{revtex}

\begin{document}
\title{Scaling solutions in general non-minimal coupling theories}
\author{Luca Amendola}
\address{Osservatorio Astronomico di Roma, Viale del Parco Mellini 84\\
00136 Roma, Italy}
\date{\today }
\maketitle

\begin{abstract}
A class of generalized non-minimal coupling theories is investigated, in
search of scaling attractors able to provide an accelerated expansion at the
present time. Solutions are found in the strong coupling regime and when the
coupling function and the potential verify a simple relation. In such cases,
which include power law and exponential functions, the dynamics is
independent of the exact form of the coupling and the potential. The
constraint from the time variability of $G$, however, limits the fraction of
energy in the scalar field to less than 4\% of the total energy density, and
excludes accelerated solutions at the present.
\end{abstract}


\section{Introduction}

A great deal of effort has been devoted in recent times to the dynamics of
scalar fields in the radiation and matter dominated era. The motivations are
manyfold: first, several theories of fundamental physics predict the
existence of scalar fields \cite{fri}\cite{fer}\cite{wet}; second, a slowly
rolling scalar field may mimic the behavior of a cosmological constant at
the present time, in agreement with a popular model of structure formation
and with the observation of an accelerated space expansion \cite{fri}\cite
{zla}\cite{cal}; third, the scalar field may alleviate the constraints on a
true cosmological constant \cite{cob}; fourth, the additional source of
fluctuations produced by the scalar field may give new observable effects on
the cosmic microwave background and on the structure formation \cite{fer} 
\cite{perr}\cite{via}.

So far, most work focused on fields with minimal coupling to gravity \cite
{fri}\cite{fer}\cite{zla} \cite{cal}\cite{lid} \cite{cop}. In this case, the
dynamics in a homogeneous and isotropic space-time is completely determined
once one specifies the matter fluid equation of state and the field
potential. For the former, the obvious choices of interest are the equation
of state of a relativistic fluid and of a pressureless one. For the latter,
although there are no observations or fundamental principles to guide our
investigation, potentials like power-laws, exponential and a handful of
other cases have been selected, basing either on simplicity or on some
particle physics model. Among the infinite solutions of the system of
equations, the attractor solutions are of course of the greatest interest.
Among the attractors, those which have a power-law behavior, denoted also as
scaling solutions, are particularly simple to find and to study.
Consequently, the study of the scalar field dynamics has focussed on the
search of {\it scaling attractors}. To be interesting for cosmological
purposes, these attractor solutions must also lead to a energy density in
the scalar field which is a non-negligible fraction of the total energy
density. Finally, if we want to explain recent observations of the
large-scale geometry of the space-time \cite{per}\cite{rei}, the scalar
factor has to be accelerated at the present.

In minimal coupling theories the Lagrangian is the sum of the
Einstein-Hilbert gravity Lagrangian and of the scalar field sector. The
non-minimal coupling (NMC) adds a new term which, in its simplest form, may
be written as (for a more general form that includes derivatives see Ref. 
\cite{ame}) 
\begin{equation}
f(\phi )R.
\end{equation}
For instance, Refs. \cite{acc}\cite{ame90}\cite{fut}\cite{kas} adopted $ 
f\sim \phi ^2$ discussing the model in the context of inflation. In Ref. 
\cite{uza} several attractor scaling solutions in the matter dominated
regime with power-law and exponential potential have been found. Other forms
of $f(\phi )$ have been considered (see e.g. \cite{ste}).

A common feature of all these investigations, perhaps obviously, is the
choice of specific potentials and coupling functions. The purpose of this
paper is to show that it is possible to find attractor solutions in NMC
models in which both the coupling $f(\phi )$ and the potential $V(\phi )$
are left unspecified, and only their relation matters. In other words, we
will find a class of models in which the dynamics of the system is
independent of the coupling and of the potential, and depends only on their
relation. In particular, we will find attractor solutions for all models for
which we can write \cite{bel} 
\begin{equation}
V(\phi )=Af(\phi )^M.  \label{condp}
\end{equation}
This relation holds, for instance, when both $V$ and $f$ are power-law, or
exponential, but is also valid for much more complicated functions, like
products of power law and exponential. In the limit of strong coupling, the
dynamics of the cosmological solution will be shown to depend essentially
only on $M$ and on the fluid matter equation of state.

After performing a conformal rescaling of the metric, the NMC system is
written as a scalar field in pure General Relativity with an exponential
potential and an extra coupling to the ordinary matter (see e.g. \cite{wet95} 
, in which however only the case $f\sim \phi ^2$ has been discussed). This
system shows a surprisingly rich phase space structure, with four different
attractors. Two of these are qualitatively similar to the attractors found
in the system without extra coupling. The other two however are new, and
have not been previously identified. Although we derived this system from a
class of NMC theories, we remark that it is interesting on its own, and many
cosmological properties of its trajectories have yet to be worked out. Here
we study it mainly to constrain the NMC model, and find that the constraint
on the variability of the gravitational constant rules out this class of
models as explanation for the accelerated expansion rate of the Universe.

In the next section we work out the field equations. In Section 3 we find
and discuss the attractor solutions, in Section 4 we discuss their
cosmological properties, and in the final Section we draw the conclusion and
point to new developments.

\section{Field equations}

Consider the Lagrangian of a NMC scalar field plus a perfect fluid matter
component ($\kappa ^2\equiv 8\pi M_p^{-2}$) 
\begin{eqnarray}
L_{tot} &=&L(\phi ,R)+2\kappa ^2L_\phi +2\kappa ^2L_{matter}, \\
L(\phi ,R) &=&-f(\phi )R, \\
L_\phi &=&\frac 12\phi _{,\mu }\phi ^{,\mu }-V(\phi ).
\end{eqnarray}
Contrary to the usual notation, we found convenient to include into $f(\phi
) $ the constant that produces the Einstein-Hilbert term, so that our $f$ is 
$1+\kappa ^2\xi f(\phi )$ in the notation of, e.g., Ref. \cite{acc}\cite
{ame90}\cite{fut}\cite{kas}). We will always assume $f>0$, since it acts as
an effective gravitational constant, 
\begin{equation}
G_{eff}=(2\kappa ^2f)^{-1}.
\end{equation}

The Einstein equations are 
\begin{equation}
G_{\mu \nu }=L_{,R}^{-1}\left[ \frac 12g_{\mu \nu }(L-L_{,R}R)-g_{\mu \nu
}\Box L_{,R}+(L_{,R})_{;\mu \nu }+\kappa ^2T_{\mu \nu \left( \phi \right)
}+\kappa ^2T_{\mu \nu \left( m\right) }\right] ,
\end{equation}
where $L_{,R}$ denotes here $dL/dR$, and where the scalar field
energy-momentum tensor is 
\begin{equation}
T_{\mu \nu \left( \phi \right) }=\phi _{,\mu }\phi _{,\nu }-\frac 12g_{\mu
\upsilon }\phi _{,\alpha }\phi ^{,\alpha }+g_{\mu \upsilon }V(\phi ),
\end{equation}
and the fluid tensor is 
\begin{equation}
T_{\mu \nu \left( m\right) }=(\rho +p)u_\mu u_\nu -g_{\mu \nu }p.
\end{equation}

Now, under the conformal transformation 
\begin{equation}
\widetilde{g}_{\mu \upsilon }=e^{2\omega }g_{\mu \upsilon },
\end{equation}
the following transformations (see e .g. \cite{bar}\cite{sch}\cite{mae})
occur: the kinetic term 
\begin{equation}
K_{\mu \nu \left( \phi \right) }=\phi _{,\mu }\phi _{,\nu }-\frac 12g_{\mu
\upsilon }\phi _{,\alpha }\phi ^{,\alpha },
\end{equation}
remains invaried ($K_{\mu \nu }=\widetilde{K}_{\mu \nu }$); the potential
term $g_{\mu \upsilon }V(\phi )$ becomes $e^{-2\omega }\widetilde{g}_{\mu
\upsilon }V(\phi )$; and the perfect fluid tensor becomes 
\begin{equation}
T_{\mu \nu (m)}=e^{-2\omega }\widetilde{T}_{\mu \nu (m)}.
\end{equation}
Putting 
\begin{equation}
2\omega =\log f,
\end{equation}
it follows that the equations in the rescaled metric (sometimes called
Einstein frame, while the old metric is the Jordan frame) are 
\begin{equation}
\widetilde{G}_{\mu \nu }=\kappa ^2\left[ F^2(\phi )\widetilde{K}_{\mu \nu
\left( \phi \right) }+\widetilde{g}_{\mu \upsilon }e^{-4\omega }V(\phi
)+e^{-4\omega }\widetilde{T}_{\mu \nu \left( m\right) }\right] ,
\end{equation}
where 
\begin{equation}
F^2(\phi )=\frac 1f+\left( \frac{f^{^{\prime }}}{cf}\right) ^2,  \label{fphi}
\end{equation}
where $c^2=2\kappa ^2/3$ and where the prime denotes derivation with respect
to $\phi $. We can then define a new canonical field 
\begin{equation}
\psi \equiv \int d\phi F(\phi ),
\end{equation}
a new potential 
\begin{equation}
U(\psi )\equiv \frac{V(\phi )}{f(\phi )^2},
\end{equation}
and a new matter tensor 
\begin{equation}
\widetilde{T}_{\mu \nu \left( m\right) }^{*}\equiv e^{-4\omega }\widetilde{T} 
_{\mu \nu \left( m\right) }.
\end{equation}
Finally, all these definitions lead to the canonical equations in the new
metric $\widetilde{g}_{\mu \nu }$ 
\begin{equation}
\widetilde{G}_{\mu \nu }=\kappa ^2\left[ \widetilde{T}_{\mu \nu \left( \psi
\right) }+\widetilde{T}_{\mu \nu \left( m\right) }^{*}\right] .
\end{equation}
The new matter energy-momentum tensor can be written as 
\begin{equation}
\widetilde{T}_{\mu \left( m\right) }^{*\nu }=diag(\rho e^{-4\omega
},-pe^{-4\omega },-pe^{-4\omega },-pe^{-4\omega })=diag(\rho
^{*},-p^{*},-p^{*},-p^{*}).
\end{equation}
As a last step, we rewrite the new metric in the Friedmannian form 
\begin{equation}
\widetilde{g}_{\mu \nu }=diag(1,-\widetilde{a}^2-\widetilde{a}^2,-\widetilde{ 
a}^2),
\end{equation}
where the old time and the old scale factor are 
\begin{equation}
t=\int e^{-\omega (\tilde{t})}d\widetilde{t},  \label{timerescal}
\end{equation}
and 
\begin{equation}
a=e^{-\omega (\widetilde{t})}\widetilde{a}.  \label{arescal}
\end{equation}
The equation of motion for the fields are obtained as the covariant
conservation laws of the energy tensors. In the old frame they read 
\begin{eqnarray}
\Box \phi +V^{\prime }+f^{\prime }R/2\kappa ^2 &=&0,  \nonumber \\
T_{\mu \nu (\phi )}^{;\mu } &=&0.
\end{eqnarray}
The transformation to the new frame is performed according to the rules 
\begin{eqnarray}
R &=&e^{2\omega }(\widetilde{R}-6\widetilde{g}^{\alpha \beta }\omega
_{,\alpha }\omega _{,\beta }+6\widetilde{\Box }\omega ),  \nonumber \\
\Box &=&e^{2\omega }(\widetilde{\Box }-2\widetilde{g}^{\alpha \beta }\omega
_{,\alpha }\nabla _\beta ).
\end{eqnarray}
From now on, we omit all the tilde, until we return to the original
quantities. Finally, the full set of equations in the Friedmann metric read: 
\begin{eqnarray}
\ddot{\psi}+3H\dot{\psi}+U_{,\psi } &=&\frac 12W_{,\psi }(\rho ^{*}-3p^{*}),
\label{sys1} \\
\dot{\rho}^{*}+3H\left( \rho ^{*}+p^{*}\right) &=&-\frac 12W_{,\psi }\dot{ 
\psi}(\rho ^{*}-3p^{*}), \\
3H^2 &=&\kappa ^2\left( \rho ^{*}+\frac 12\dot{\psi}^2+U\right) , \\
-2\dot{H} &=&\kappa ^2\left( \rho ^{*}+p^{*}+\dot{\psi}^2\right) ,
\label{sys4}
\end{eqnarray}
where 
\begin{equation}
W=\log f(\phi ),\quad W_{,\psi }=\frac{f^{\prime }}{fF}.
\end{equation}
As already remarked in the Introduction, the system (\ref{sys1}-\ref{sys4}),
here derived from a NMC model, is interesting on its own. Indeed, we can
regard either the Jordan or the Einstein frame as the physical one. In the
former case, we have to express the solutions of the above system back in
the original frame, and study its cosmological consequences in the original
frame, as we will do below. In the latter case, the solutions of the system
are the physical solutions, and their properties can be directly compared to
observations. In particular, the constraints from the variability of $G$ ,
which we will find to limit heavily the cosmological viability of our
solutions, apply only assuming the physical frame to be the original Jordan
one.

\section{Solutions}

The full dynamics of the system (\ref{sys1}-\ref{sys4}) is specified by the
potential $U$ and by the equation of state $p=(w-1)\rho $. In the following
we consider only $0\leq w\leq 2$. To write down the potential $U(\psi )$, we
have first to find the relation between $\psi $ and $\phi $. This is where
the possibility of a dynamics independent of the potential and of the
coupling function arises. In fact, if we assume that \cite{bel} 
\begin{equation}
f^{\prime 2}\gg c^2f,  \label{condf}
\end{equation}
then we can simplify Eq. (\ref{fphi}): 
\begin{equation}
F^2(\phi )=\left( \frac{f^{^{\prime }}}{cf}\right) ^2.  \label{fdef}
\end{equation}
It follows 
\begin{equation}
c\psi =\int \frac{df}f=\log f,  \label{psif}
\end{equation}
where the integration constant can be absorbed into a redefinition of $\psi $
. It follows that the conformal function $\omega $ equals $c\psi /2$.
Therefore, once we have the dynamics of $\psi $ in the transformed metric,
we can write down the solution in terms of the original metric without
having to specify $f(\phi ),$ provided we express also the potential $V(\phi
)$ as a function of $f(\phi )$. With the assumption (\ref{condf}) we get $ 
W_{,\psi }=c$ in the system (\ref{sys1}-\ref{sys4}), so that putting 
\[
\beta =4-3w, 
\]
the first two equations become 
\begin{eqnarray}
\ddot{\psi}+3H\dot{\psi}+U_{,\psi } &=&\frac 12c\beta \rho ^{*},  \nonumber
\label{wet} \\
\dot{\rho}^{*}+3Hw\rho ^{*} &=&-\frac 12c\beta \dot{\psi}\rho ^{*}.
\label{wet}
\end{eqnarray}

The condition (\ref{condf}) holds true in several cases. For instance, it is
verified for large $\phi $ by any function $f(\phi )$ which grows faster
than quadratically, that is $\lim_{\phi \rightarrow \infty }f(\phi )/\phi
^2\rightarrow \infty $. In the often-studied quadratic case, $f=1+\kappa
^2\xi \phi ^2$, for large $\phi $ we can put $f(\phi )=\kappa ^2\xi \phi ^2$
. Then, instead of Eq. (\ref{condf}), one has $f^{\prime 2}=4\kappa ^2\xi f$
, and all that changes is that in Eq. (\ref{fdef}) and Eq. (\ref{psif}) $c^2 
$ is replaced by $c^2/(1+1/6\xi )$. In this case, all the results found
below become {\it exact}. The weak coupling limit in which $\kappa ^2\xi
\phi ^2\ll 1$, i.e. $\xi \ll (\kappa ^2\phi ^2)^{-1}$, on the other hand, is
excluded in the present analysis. We could then label our case as the {\it  
strong coupling} limit. In fact, it is easily seen that it corresponds to
the limit in which the Lagrangian can be approximated as $-f(\phi )R-2\kappa
^2V(\phi )$, neglecting the kinetic term $\frac 12\phi _{,\mu }\phi ^{,\mu }$
. Notice however that this does not imply that the scalar field kinetic
terms in the field equations are negligible, because the non-minimal
coupling itself introduces other kinetic terms.

Now, as anticipated, suppose we can write $V(\phi )=Af(\phi )^M$ .The
potential becomes then 
\begin{equation}
U(\psi )=\frac{Af(\phi )^M}{f(\phi )^2}=Ae^{\sqrt{2/3}\mu \kappa \psi },
\end{equation}
where 
\begin{equation}
\quad \mu \equiv M-2.
\end{equation}
Therefore, the potential can be written as an exponential, whatever the
shape of $V$ and of $f$, provided that the condition (\ref{condf}) and the
relation (\ref{condp}) are fulfilled. The sign of $\mu $ selects the
direction in which the field $\psi $, and thus the variable $f$, rolls. If $ 
\mu >0$, $\psi $ rolls toward $-\infty $, so that $f\rightarrow 0$, and the
effective gravitational constant $G_{eff}$ increases with time. In the
opposite case, $\mu <0$, we have that $G_{eff}$ decreases in the future. We
emphasize that if $f(\phi )$ is quadratic, then all results below remain
valid provided $\mu $ is replaced by $\mu _q=\mu /(1+1/6\xi )^{1/2}$ and $ 
\beta $ by $\beta _q=\beta /(1+1/6\xi )^{1/2}.$

The scalar field dynamics in NMC theories is then reduced to the scalar
field dynamics in pure general relativity with an exponential potential and
with a scalar field/matter coupling. In the radiation case in which $w=4/3$,
we have $\beta =0$, and the source terms decouple. The decoupling occurs
also when we can neglect the matter energy density $\Omega _{\rho
^{*}}=\kappa ^2\rho ^{*}/3H^2$ with respect to the scalar field energy
density $\Omega _\psi $. In these cases, the problem is identical to that
already solved in, e.g., Ref. \cite{fer}\cite{lid}\cite{cop}\cite{rat}. The
case $\beta \neq 0$ has been already discussed by Wetterich in \cite{wet95},
where some of its attractors have been identified. Here we extend the
analysis to the full classification of critical points and attractors
(finding two new attractors) and express the solutions in terms of the old
frame. We keep $\beta $ as an independent parameter as long as possible, and
proceed to replacing it by $4-3w$ only in the graphics, in order to narrow
the parameter space to two dimensions, namely $w$ and $\mu $. The formulas
apply however to the more general case, unless otherwise specified.

Following Copeland et al. \cite{cop} we define 
\begin{equation}
x=\frac{\kappa \dot{\psi}}{\sqrt{6}H},\quad y=\frac{\kappa \sqrt{U}}{\sqrt{3} 
H},
\end{equation}
and introduce the independent variable $\alpha =\log a(t)$. Notice that $x^2$
and $y^2$ give the fraction of total energy density carried by the scalar
field kinetic and potential energy, respectively . Then, we can rewrite the
system (\ref{sys1}-\ref{sys4}) as 
\begin{eqnarray}
x^{\prime } &=&-3x+3x\left[ x^2+\frac 12w(1-x^2-y^2)\right] -\mu y^2+\frac 12 
\beta (1-x^2-y^2),  \nonumber \\
y^{\prime } &=&\mu xy+3y\left[ x^2+\frac 12w(1-x^2-y^2)\right] .
\label{syst2}
\end{eqnarray}
where the prime is here $d/d\alpha $. The system is invariant under the
change of sign of $y$ and of $\alpha $. Since it is also limited by the
condition $\rho ^{*}>0$ to the circle $x^2+y^2\leq 1$, we may study only the
unitary semicircle of positive $y$. The critical points, those that verify $ 
x^{\prime }=y^{\prime }=0$, are scaling solutions, on which the scalar field
equation of state is 
\begin{equation}
w_\psi =\frac{2x^2}{x^2+y^2}=const,
\end{equation}
the scalar field total energy density is $\Omega _\psi =x^2+y^2$, and the
scale factor is 
\begin{equation}
a\sim t^p,\quad p=\frac 2{3w}\left[ \frac w{w+\Omega _\psi (w_\psi -w)} 
\right]
\end{equation}
(the slope $p$ is not to be confused with the pressure).

Copeland et al. \cite{cop} have shown that the system (\ref{syst2}) with $ 
\beta =0$ and an exponential potential has up to five critical points, that
can be classified according to the dominant energy density: one dominated by
the scalar field total energy density (let us label this point as solution $ 
a $ and refer to its coordinates as $x_a,y_a$ ), one in which the fractions
of energy density in the matter and in the field are both non-zero (labelled 
$b$), one dominated by the matter field ($c$), and finally two dominated by
the kinetic energy of the scalar field, of which one at $x=-1$ ($d$) and one
at $x=+1$ ($e$).

The critical points on which the matter field becomes negligible reduce to
the $\beta =0$ case: therefore, the solutions $a,$ $d$, and $e$ remain the
same also for $\beta \neq 0.$ The points $b$ and $c$ are instead modified.
The solution $c$ is no longer matter dominated: rather, the scalar field
kinetic energy and the matter energy take up a constant fraction of the
total energy. In MDE, the scalar field kinetic energy amounts to $\Omega
_\psi =1/9.$ The critical points in the general case $\beta \neq 0$ are
listed in Tab. I, where we put $g(\beta ,w,\mu )\equiv \beta ^2+2\beta \mu
+18w.$ 
\[
\begin{tabular}{|c|c|c|c|c|c|}
\hline
& $x$ & $y$ & $\Omega _\psi $ & $p$ & $w_\psi $ \\ \hline
$a$ & $-\mu /3$ & $\left( 1-x_a^2\right) ^{1/2}$ & $1$ & $3/\mu ^2$ & $2\mu
^2/9$ \\ \hline
$b$ & $-\frac{3w}{2\mu +\beta }$ & $-x_b\left( \frac g{9w^2}-1\right) ^{1/2}$
& $\frac g{\left( \beta +2\mu \right) ^2}$ & $\frac 2{3w}\left( 1+\frac \beta
{2\mu }\right) $ & $\frac{18w^2}g$ \\ \hline
$c$ & $\frac \beta {6-3w}$ & $0$ & $\left( \frac \beta {6-3w}\right) ^2$ & $ 
\frac{6(2-w)}{\beta ^2+9(2-w)w}$ & $2$ \\ \hline
$d$ & $-1$ & $0$ & $1$ & $1/3$ & $2$ \\ \hline
$e$ & $+1$ & $0$ & $1$ & $1/3$ & 2 \\ \hline
\multicolumn{6}{|c|}{Tab. I} \\ \hline
\end{tabular}
\]

Although the number and position of the critical points is affected only
quantitatively by the extra coupling, their stability properties are
modified in a more radical way. In particular, while for $\beta =0$ only the
points $a$ and $b$ can be attractors, here we show that also $c$ and $d$ may
be stable. Only the point $e$ remains always unstable.

The stability analysis is performed as usual by linearization around the
critical points. The parametric regions in which the real part of both
eigenvalues of the linearization matrix is negative are regions of
stability. To simplify the discussion, we only consider the crucial property
of stability versus instability, paying no attention to the topography of
the critical point (whether it is a knot, spiral, or saddle). In the
following, we say that an attractor exists if it lies in the region $0\leq
x^2+y^2\leq 1$. The parameter spaces are plotted in Fig. 1.

{\bf Point }$a$ .

The solution $a$ exists for $\mu ^2<9$, and is an attractor only for $\mu
_{-}<\mu <\mu _{+}$ where 
\begin{equation}
\mu _{\pm }=\frac 14\left( -\beta \pm \sqrt{\beta ^2+72w}\right) ,
\end{equation}
(e.g., $\mu _{-}=-2.39$ and $\mu _{+}=1.89$ for $w=1$). On this attractor we
have $w_\psi =2\mu ^2/9$ and 
\begin{equation}
p_a=3/\mu ^2,
\end{equation}
inflationary if $|\mu |<\sqrt{3}$.

{\bf Point }$b$ .

The attractor $b$ exists and is stable in the region delimited by $\mu <\mu
_{-}$ and $\mu >\mu _{+}$ and the two branches of the curve 
\begin{equation}
\mu _0=-\frac 1{2\beta }\left( \beta ^2+18w-9w^2\right) .
\end{equation}
The scale factor slope on the attractor is 
\begin{equation}
p_b=\frac 2{3w}\left( 1+\frac \beta {2\mu }\right) .
\end{equation}
and, for $\beta =4-3w$, is inflationary within the two branches of the curve 
\begin{equation}
\mu _i=\frac{4-3w}{3w-2}.
\end{equation}
It is remarkable that the inflationary region for the point $b$ includes
values smaller than $w\approx 0.91$, and therefore {\it excludes} the MDE
equation of state $w=1$. This conclusion is not changed by replacing $\mu $
and $\beta $ with their counterparts $\mu _q$ and $\beta _q$ in the case of
a quadratic coupling $f(\phi )$.

{\bf Point }$c$ .

This point exists for $w<5/3$, and is stable below the lower branch and
above the upper branch of $\mu _0$ $.$ The slope is 
\[
p_c=\frac{6(2-w)}{\beta ^2+9(2-w)w}, 
\]
and it is never accelerated if $\beta =4-3w$. The point $c$ shares with $b$
the property that matter and scalar field have both a non-vanishing fraction
of the energy density.

{\bf Point }$d$ .

This point exists for all values of the parameters, and if $\beta =4-3w$ is
stable for $w>5/3$ and $\mu >3$. Its slope is always $p_d=1/3$.

{\bf Point }$e.$

This point exists and is unstable for all values of the parameters if $\beta
=4-3w$.

The complex structure of the parameter space is summarized in Fig. 2. Notice
that 1), for each value of the parameters $w,\mu $ there is one and only one
attractor; 2) for $w=1$ the points $a,b$ or $c$ can be stable, depending on $ 
\mu $; 3) these solutions are inflationary in the shaded region; 4) only the
point $a$ can be accelerated for $w=1$ or larger. In Fig. 3 we present four
phase spaces displaying in turn the four possible attractors. The parameters
correspond to the points marked with stars in Fig. 2. As already remarked,
attractors $c$ and $d$ have not been previously noticed. Also, it is
important to remark that the attractors are not only locally stable, but
extend their basin of attraction to all of the phase space. That is, {\it any 
} possible initial condition lead to the attractor.

\section{Back to the Jordan frame}

Here we leave the dynamical analysis of the system in the rescaled frame and
get back to the original one. What the attractors look like in the Jordan
frame?

Reintroducing the tildes, we have along the attractors $a$ and $b$ (from now
on, quantities without tildes are expressed in the original metric) 
\begin{equation}
c\psi =-\frac 2\mu \log |\widetilde{t}/\tau _{a,b}|,  \label{tab}
\end{equation}
(for $\mu \neq 0$) where 
\begin{equation}
\tau _{a,b}^{-1}=\sqrt{2A}\mu c\frac{x_{a,b}}{y_{a,b}}.
\end{equation}
On the attractors $c$ and $d$ , for which $y=0$, $\psi \rightarrow \infty $,
and the conformal transformation cannot be performed. Since $\psi $ is
proportional to $\log f$, the attractors $c$ and $d$ lead to an effective
gravitational constant that is either zero or infinite and are therefore to
be rejected as possible solutions in the Jordan frame. Of course,
trajectories that have not already reached the attractor cannot be excluded,
but these are not scaling solutions, and will not be further considered in
this paper.

Form Eq. (\ref{tab}) it follows (neglecting the subscripts) 
\begin{equation}
e^{2\omega }=(\widetilde{t}/\tau )^{-2/\mu }.
\end{equation}
From the latter expression we can evaluate the relation between the old and
new time and scale factor, given by Eq. (\ref{timerescal}) and (\ref{arescal} 
). We obtain (for $\mu \neq 0,-1$) 
\begin{eqnarray}
\widetilde{t} &\sim &t^{\frac \mu {1+\mu }},  \nonumber \\
\widetilde{a} &\sim &t^{-1/(1+\mu )}a(t).
\end{eqnarray}
As can be seen, for $\mu \rightarrow \pm \infty $ the old and new metric
coincide; in this limit the scalar field vanishes on the attractor, and the
system reduces to the pure perfect fluid Friedmann case.

It follows that in the original variables the scale factor is again a power
law 
\begin{equation}
a\sim t^{p^{\prime }},\quad p^{\prime }=\frac{1+\mu p}{1+\mu }.
\end{equation}
On the attractor $a$, $p_a=3/\mu ^2,$ which is inflationary (both in the
original and in the rescaled frame) if $\mu ^2<3$, that is $2-\sqrt{3}<M<2+ 
\sqrt{3}$ , we have 
\begin{equation}
p_a^{\prime }=\frac{\mu +3}{\mu (1+\mu )}.
\end{equation}
Consider now some special cases. If $0>\mu >-1$ the scale factor follows a 
{\it pole-like} inflation, $a\sim (t_0-t)^{p_a^{\prime }}$ with negative
exponent. For $\mu =0$ ,(i.e. $V\sim f^2$) the old and new metric coincide
(up to a constant), the field freezes to a constant and its energy drives a
deSitter expansion. The system reduces asymptotically to pure general
relativity with a cosmological constant. Finally, for $\mu =-1$ (i.e. $V\sim
f$), the scale factor is power law accelerated in the new frame, but maps
again to a deSitter expansion in the original frame. If $f$ is quadratic in $ 
\phi $, then the inflationary condition on the solution $a$ reads 
\begin{equation}
\mu ^2<3(1+1/6\xi ).  \label{acc}
\end{equation}
On the attractor $b$ , on the other hand, putting $p_b=2/(3w^{\prime })$
with $w^{\prime }=w\left( 1+\beta /2\mu \right) ^{-1}$ we obtain 
\begin{equation}
p_b^{\prime }=\frac{3w^{\prime }+2\mu }{3w^{\prime }(1+\mu )}.
\end{equation}
Notice that $p_b^{\prime }\rightarrow 2/3w$ for $\mu \rightarrow \pm \infty $
, as expected. Since the property of being accelerated is conformally
invariant (for positive definite conformal factors), going back to the old
frame do not change qualitatively the attractors found so far. Also, it is
not difficult to check that $\Omega _{\rho ^{*}}=\kappa ^2\rho ^{*}/3 
\widetilde{H}^2$ and $\Omega _\psi $ are invariant under conformal
transformation, so that $\Omega _\psi =\Omega _\phi $ .

It can be shown that the choice $V\sim f^M$ is the only one that allows
scaling attractors in both the old and the new metric. Other choices are
possible that allow scaling solutions either in the old or in the new
metric: for instance, $V\sim f^2\left( \log f\right) ^M$ gives scaling
attractors in the new metric but not in the old one.

\section{Cosmological properties}

Once we have the analytical expression of the attractors, we must consider
whether they are viable as cosmological solutions. The attractor solution $a$
is inflationary (accelerated) and the scalar field is asymptotically the
dominating component. As such, it may match the observations of an
accelerated expansion; for instance, the value $w_\psi \approx 0.4$
suggested in Ref. \cite{tur} implies 
\[
M\approx 0.7\text{ or }3.3. 
\]
On the other hand, since $\Omega _\phi \rightarrow 1$, in order to allow for
a substantial fraction in the ordinary matter component at the present, the
attractor has not to be already reached.

The solution $b$ has some drawbacks. First, is not accelerated at all for $ 
w=1$; second, the constraints from nucleosynthesis do not allow a large
fraction of energy density in the scalar field, so that it cannot provide
closure energy. However, as argued in \cite{fer}, models which reach this
attractor compare favorably with observations of large scale structure, and
may have a simple interpretation in terms of fundamental physics.

Both solutions $a$ and $b$ are heavily constrained by the upper limits on
the variability of the gravitational constant. We have 
\begin{equation}
|\dot{G}/G|=|\dot{f}/f|=\frac 2{|1+\mu |}\frac 1t.
\end{equation}
Comparing with the observational constraint $|\dot{G}/G|<a10^{-10}yr^{-1}$,
and assuming $t\approx 10$ Gyr, we obtain the condition 
\begin{equation}
\mu >\frac 2a-1.  \label{cn}
\end{equation}
Current constraints (see e.g. \cite{gue} ) give $a\approx 0.1$ or smaller.
This implies $\mu >20$, too large for the attractor $a$ to exist. A similar
problem arises if $f$ is quadratic. Along the attractor $b$, the energy
density in the scalar field is a constant fraction of the total energy. In
MDE (and for $\beta =4-3w=1$) this is 
\begin{equation}
\Omega _\phi =\frac{19+2\mu }{\left( 1+2\mu \right) ^2}.
\end{equation}
The constraint (\ref{cn}) gives 
\begin{equation}
\Omega _\phi \leq 0.035,
\end{equation}
which confines the scalar field contribution to that of a minor component.
This constraint is three or four times stronger than that imposed by the
nucleosynthesis \cite{fer} on a minimally coupled field.

\section{Conclusions}

In this paper we have investigated a large class of NMC models in the limit
of strong coupling with a perfect fluid matter component, searching for
attractors that might provide a decaying cosmological constant. These models
include all the cases in which the potential $V(\phi )$ is a power of the
coupling function $f(\phi )$, regardless of their functional form. We have
shown that

\begin{enumerate}
\item  The NMC system can be reduced to a scalar field with an exponential
potential, a minimal coupling to gravity, and an extra coupling to the
matter.

\item  For each pair of the parameters $w,\mu $ there is one out of four
possible scaling attractors: one, $a$, scalar field dominated and possibly
accelerated; one, $b$, decelerated if $w\geq 0.91$ and with constant ratio
of scalar field total energy to matter; one, $c$, always decelerated and
with constant ratio of scalar field kinetic energy to matter; and finally
one, $d$, also always decelerated, and dominated by the field kinetic energy.

\item  Attractors $c$ and $d$ are acceptable only in the rescaled frame; in
the original frame they lead to a gravitational constant either vanishing or
infinite.

\item  This choice $V\sim f^{M\text{ }}$ is the only choice (in the strong
coupling regime) for which there is a scaling attractor both in the original
and in the rescaled metric.

\item  The constraint on the time variability of $G$ rules out the
accelerated models, and only allows a very small fraction of the energy
density to be in the NMC scalar field.
\end{enumerate}

Clearly, this analysis is not yet conclusive. Viable solutions might exist
for which one or more of the following is true: $a$) the attractors are not
yet reached; $b$) $V$ does not equal $f^M$; $c$) the strong coupling regime
does not apply. For instance, assuming $f=1+\kappa ^2\xi \phi ^2$, and in
the limit of weak coupling , the $|\dot{G}/G|$ bound can be satisfied for
small $\xi $, and the solutions are cosmologically acceptable, although by
construction do not add much to the minimally coupled model.

\section{Acknowledgments}

I am indebted to Carlo Baccigalupi, Francesca Perrotta and Jean-Philippe
Uzan for useful discussions on the topic.

\newpage\ 

\section{Figure Caption}

Fig. 1.

Regions of existence and stability in the parameter space $w,\mu $ . In all
panels, the eigenvalues of the linearization matrix change sign across the
thick lines. The color code is as follows: white regions indicate that the
critical point does not exists; light gray regions, the point is unstable;
dark gray regions, the point is stable. The dotted lines are at $w=4/3$ and $%
w=5/3$ and are useful landmarks in the parameter space. From top to bottom,
parameter spaces of the critical points $a,b,c$ and $d$.

Fig. 2.

Regions of stability in the parameter spaces. Each region is labelled by the
critical point that is stable in that region. The gray area indicates where
the attractor is accelerated. The stars mark the values of the parameters
for which we display in the next plot the phase space.

Fig. 3.

Phase spaces for various values of the parameters, corresponding to the
points marked as stars in Fig. 2. The phase space is contained in the
positive unitary semicircle. While the phase space of attractors $a$ and $b$
are qualitatively similar to those displayed in Copeland et al. \cite{cop},
the phase space of attractors $c$ and $d$ have no counterpart for $\beta =0$.

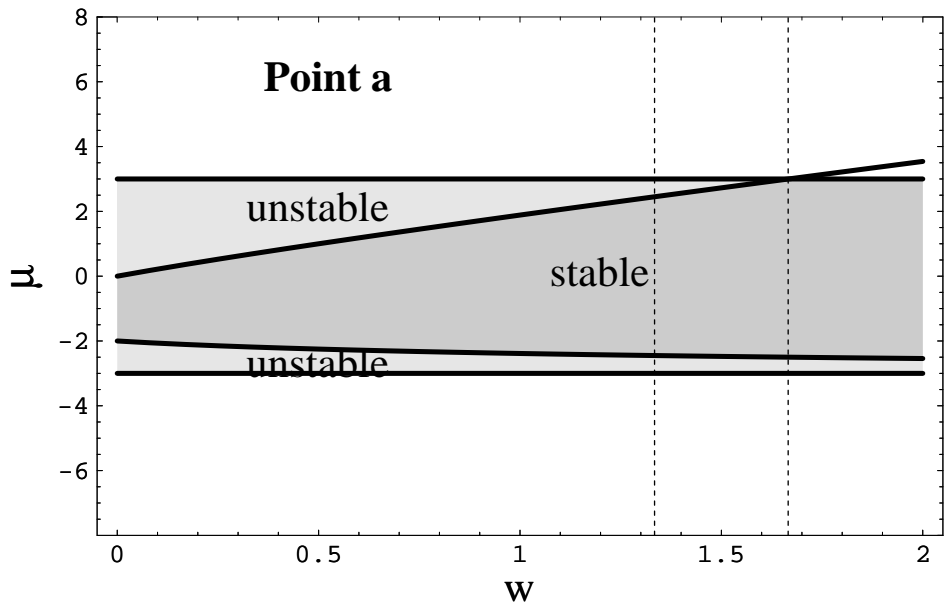
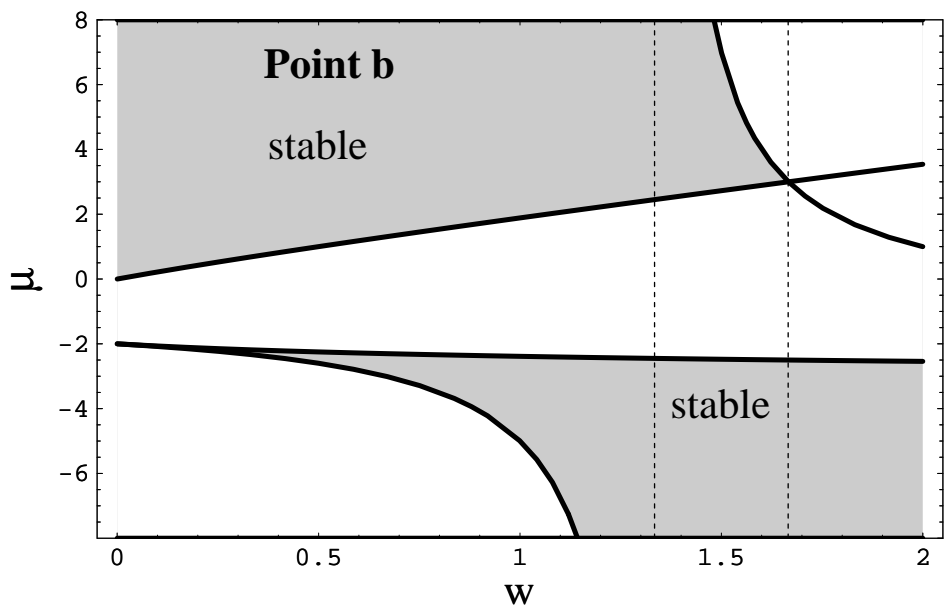
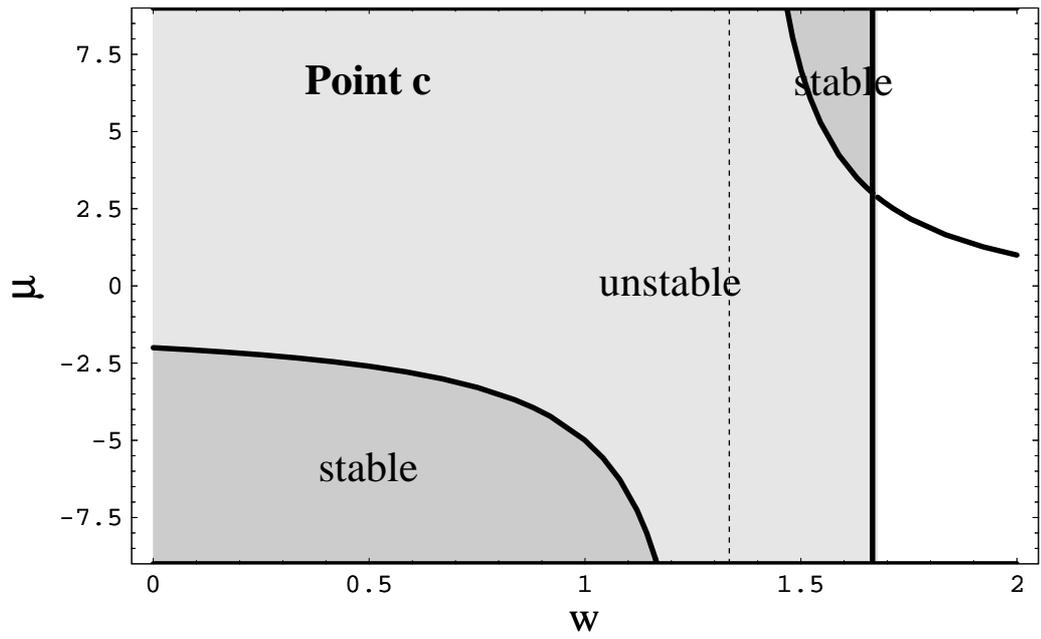
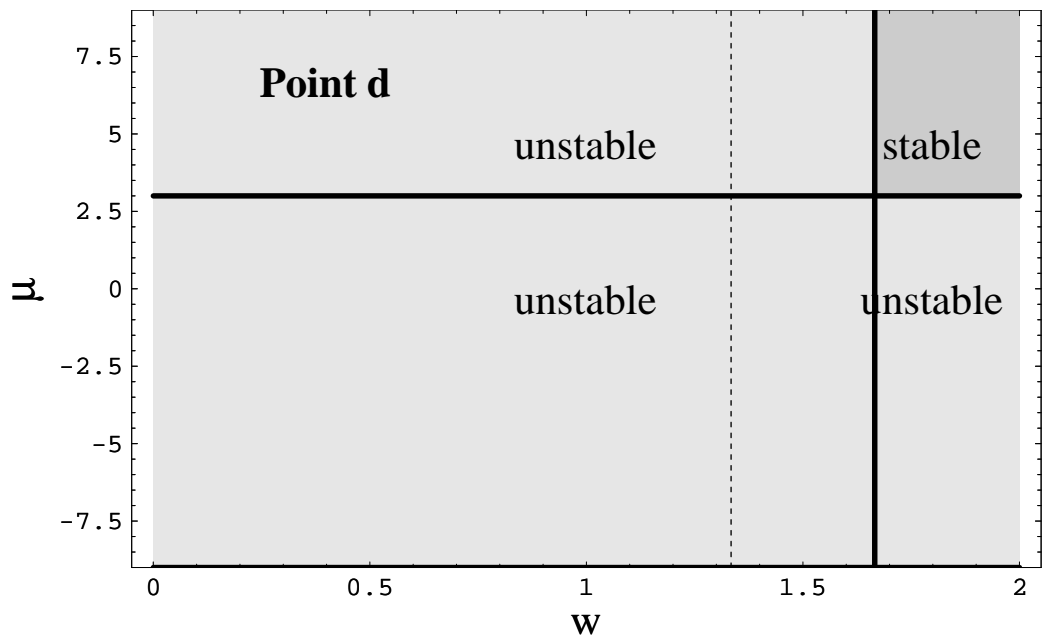
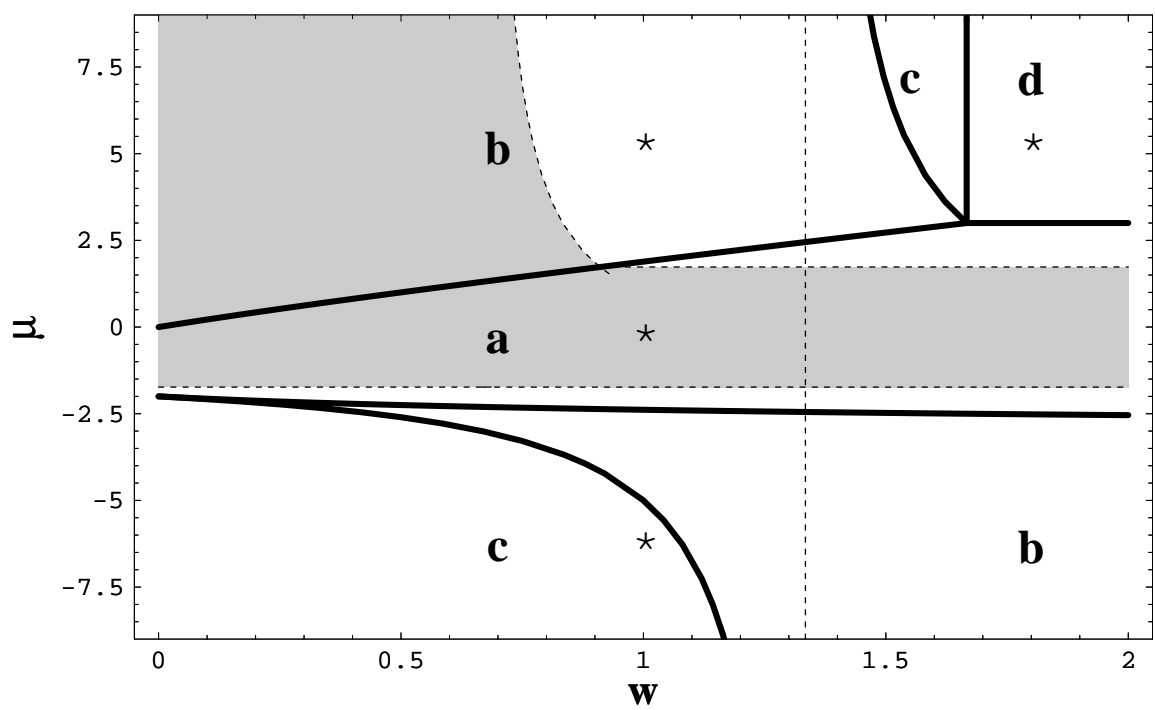
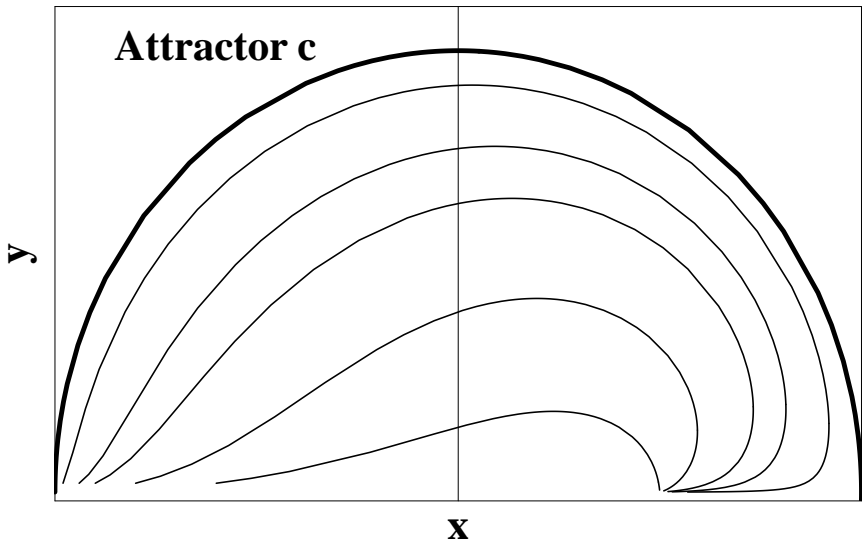
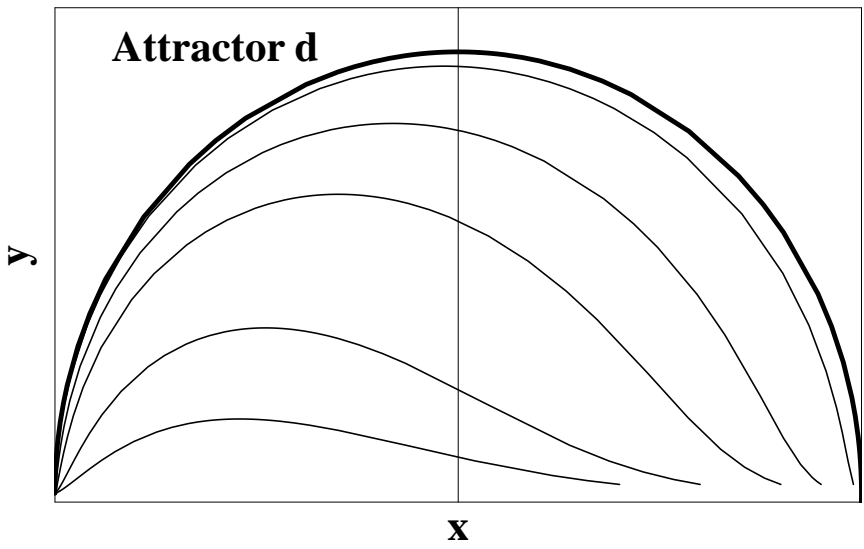
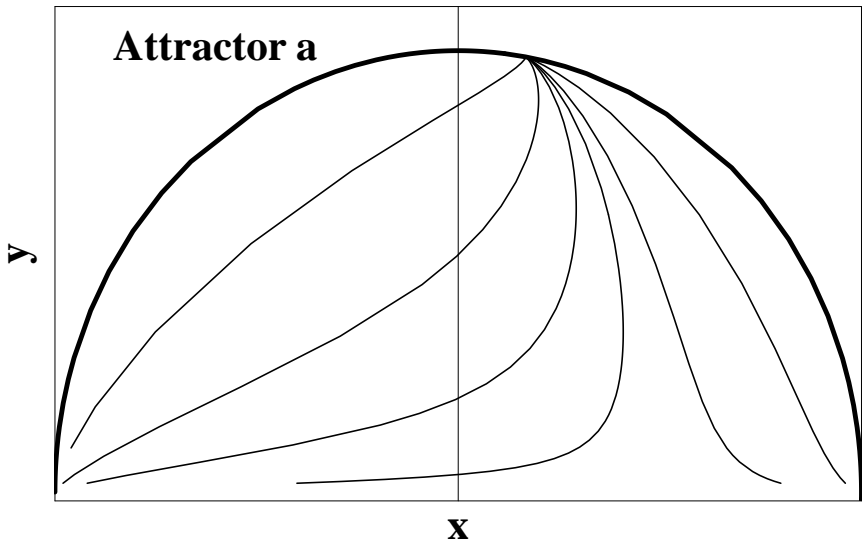
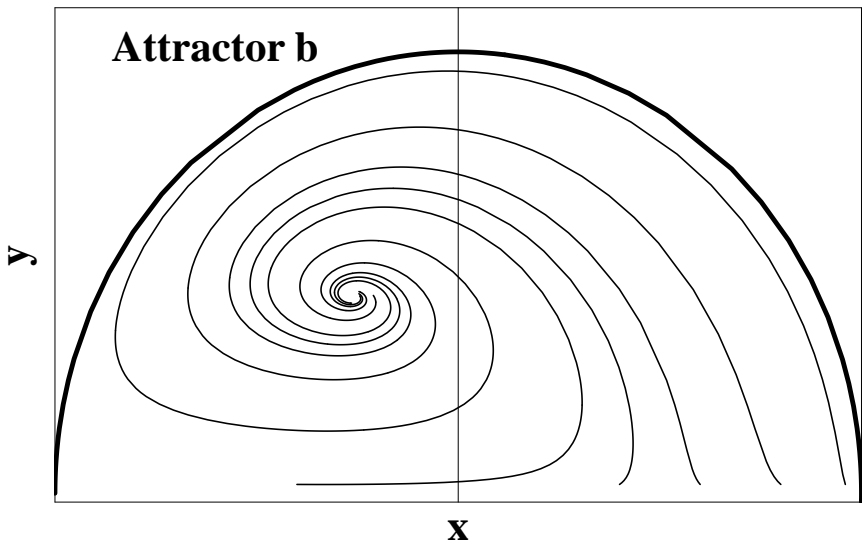
\end{document}